\documentclass[sigconf]{acmart}

\usepackage{todonotes} 

\newcommand{\Clew}{Clew}
\newcommand{\ClewMaps}{ClewMaps}
\newcommand{\InvisibleMap}{Invisible Map}
\newcommand{\PathBlazer}{NaviShare}

\AtBeginDocument{%
  \providecommand\BibTeX{{%
    \normalfont B\kern-0.5em{\scshape i\kern-0.25em b}\kern-0.8em\TeX}}}

\begin{document}

\title{Large-scale, Longitudinal, Hybrid Participatory Design Program to Create Navigation Technology for the Blind}


\author{Daeun Joyce Chung}
\authornote{Both authors contributed equally to this research.}
\affiliation{%
  \institution{Wellesley College}
  \state{MA}
  \country{USA}}
\email{dc103@wellesley.edu}

\author{Muya Guoji}
\authornotemark[1]
\affiliation{%
  \institution{Babson College}
  \state{MA}
  \country{USA}}
\email{mguoji1@babson.edu}

\author{Nina Mindel}
\affiliation{%
  \institution{Olin College of Engineering}
  \state{MA}
  \country{USA}}
\email{nmindel@olin.edu}

\author{Alexis Malkin}
\affiliation{%
  \institution{New England College of Optometry}
  \state{MA}
  \country{USA}}
\email{malkina@neco.edu}

\author{Fernando Albertorio}
\affiliation{%
  \institution{Independent}
  \state{FL}
  \country{USA}}
\email{fernando.albertorio@gmail.com}

\author{Shane Lowe}
\affiliation{%
  \institution{Independent}
  \state{KY}
  \country{USA}}
\email{shane.lowe6989@gmail.com}

\author{Chris McNally}
\affiliation{%
  \institution{Independent}
  \state{MA}
  \country{USA}}
\email{chris@imcnally.com}

\author{Casandra Xavier}
\affiliation{%
  \institution{Independent}
  \state{MA}
  \country{USA}}
\email{cxavier@olin.edu}

\author{Paul Ruvolo}
\affiliation{%
  \institution{Olin College of Engineering}
  \streetaddress{1000 Olin Way}
  \city{Needham}
  \state{MA}
  \country{USA}
  \postcode{02492}}
\email{pruvolo@olin.edu}

\renewcommand{\shortauthors}{Chung, Guoji, Mindel, Ruvolo, et al.}

\begin{abstract}
Empowering people who are blind or visually impaired (BVI) to enhance their orientation and mobility skills is critical to equalizing their access to social and economic opportunities.  To manage this crucial challenge, we employed a novel design process based on a large-scale, longitudinal, community-based structure.  Across three annual programs we engaged with the BVI community in online and in-person modes. In total, our team included 67 total BVI participatory design participants online, 11 BVI co-designers in-person, and 4 BVI program coordinators. Through this design process we built a mobile application that enables users to generate, share, and navigate maps of indoor and outdoor environments without the need to instrument each environment with beacons or fiducial markers.  We evaluated this app at a healthcare facility, and participants in the evaluation rated the app highly with respect to its design, features, and potential for positive impact on quality of life.
\end{abstract}

\ccsdesc[500]{Human-centered computing~Accessibility design and evaluation methods}

\keywords{Participatory Design, Co-design methods, Inclusive design, Assistive Technology, Mobile Navigation Application, Indoor and Outdoor Navigation for the Blind}



\maketitle

\section{Introduction}

BVI individuals face challenges in navigating unfamiliar indoor and outdoor environments \cite{wiener2010foundations, dias2015indoor, Loomis1993, Banovic2013}.  In such situations, BVI individuals typically rely on landmarks (e.g., based on information about the environment communicated to them ahead of time by an orientation and mobility instructor, O\&M) as a means of finding their way through the environment.  If such information is not available ahead of time, instead, individuals might use Global Positioning System (GPS)-based navigation technology \cite{quinones2011supporting} or ask for directions from a stranger when they arrive at the venue.  While these strategies are often helpful for navigation, GPS-based technology has limited precision outdoors (and is  unusable indoors) \cite{quinones2011supporting} while asking for help from strangers is often unreliable and erodes one's sense of independence.  As such, accurate wayfinding remains a significant concern for navigating unfamiliar indoor and outdoor environments \cite{buimer2017innovationagenda}.

With the goal of empowering BVI people to navigate more independently and easily, researchers have explored the creation of technological O\&M tools. For example, mobile applications have been developed for indoor or outdoor navigation \cite{blindsquare, Lazarillo, ahmetovic2016navcog, Goodmaps, yoon2019leveragingar}.  However, these apps often don't support both indoor and outdoor navigation, have limited precision, require extensive instrumentation of the space (e.g., with Bluetooth Beacons) \cite{kunhoth2020indoor, ahmetovic2016navcog}, or utilize expensive and specialized hardware \cite{Goodmaps} for mapping. Further, there has been a history of the development of mobility tools for BVI individuals that do not sufficiently involve members of this group in the design conversation, potentially leading to a misunderstanding of their needs in terms of functionality and interface \cite{wiener2010foundations}.

To address the need for accurate, versatile navigation technology for individuals who are blind, we developed a participatory and co-design program centering around the creation of an application called {\PathBlazer{}}.  Our work on this app grew out of an earlier app we created called {\Clew}, which was initially created in 2017.  {\Clew} supported users in recording indoor routes for their personal use.  Specifically, users could record routes by walking from point A to point B while collecting imagery and inertial data with their smartphone.  Route start and end locations were marked by anchor points, created by physically positioning a smartphone against an orienting surface (e.g., a door frame or wall).  To navigate a saved route, users would position their phone at the anchor point and the phone would provide automated guidance to their destination in the form of audio, sound, speech, and visual feedback.  At the start of this study, {\Clew} had a monthly unique user base of about 1,000 users.

Based on the premise that with a structured and inclusive design process we could improve the app, we collaborated with BVI individuals over three summer programs, hosted during the undergraduate summer research programs of our institution. Many of the BVI collaborators participated in multiple years of our programs and were from many different regions of the world.  Most BVI participants in the design programs engaged virtually using online collaboration techniques and gave feedback on the use cases and suggestions for new features on the app through an iterative design process, while BVI co-designers were part of the in-person R\&D team, which also included undergraduate students, and worked to both design and implement the app itself. BVI program coordinators organized the program and streamlined its process with the research coordinator. By utilizing a hybrid approach we were able to incorporate a variety of perspectives to define and refine the app features and use cases. Throughout the program, we continuously refined our methodology based on the qualitative and quantitative feedback from the BVI participants, addressing structural, organizational, and accessibility concerns within our virtual spaces and tools.  This design process resulted in the creation of \PathBlazer{}, which is a fully open platform for accessible indoor and outdoor wayfinding.  This app progressively links a network of destinations through recorded route segments, integrates outdoor navigation by fusing visual localization with GPS data, and employs marker-less mapping and localization for indoor positioning.

The main research contributions of this work are:  
\begin{itemize}
\item With respect to the development of technology, we designed, created, and evaluated a mobile wayfinding application for both indoor and outdoor navigation that enables users to generate, share, and navigate maps. 
\item With respect to the exploration of best practices in human-centered designed, we implemented, refined, and evaluated a large-scale, longitudinal, hybrid participatory design that lead the design and development of our mobile app.
\end{itemize}

\section{Related Work} \label{sec:relatedwork}
As our work contains contributions both in technology and design process, in this section we will situate our work within prior literature in both of these areas.

\subsection{Importance of Centering BVI Users in the Design Process}

Participatory design and co-design initiatives highlight the importance of adopting context-specific approaches to address the diverse needs of various stakeholders \cite{smith2017participatory}. By working closely with BVI individuals, researchers can \textit{``avoid importing visual assumptions and biases''} \cite{kamat2022tangibleconstructionkit}, and by emphasizing the users' involvement and long-term needs, participatory or co-design methods may avoid assistive technology abandonment \cite{phillips1993predictors, riemer2000factors}. Reconciling the diverse needs and preferences of users within the same collective is a primary challenge in participatory design \cite{laitano2017participatory}. For example, people with low vision may prefer large, high-contrast visual interfaces to use in concert with screen magnifiers while individuals who have very little functional vision may rely on technology being accessible entirely through tactile and auditory means \cite{laitano2017participatory}. Furthermore, it is important to balance functional concerns with the form that technology takes.  For example, it may be important to understand the experiences of users with disabilities who may feel self-conscious when using assistive technologies in public settings by striving for designs for social acceptability \cite{shinohara2011shadow}.  In this work, individuals from the BVI community are involved at all levels of the project, including as authors on this paper, computer scientists, co-designers, program coordinators, and prototype evaluators.

\subsection{Online vs. In-person Participatory Design} \label{sec:onlinevsinperson}
There is a robust literature of participatory, co-design or collaborative studies with BVI individuals to develop assistive technology.  Of these studies, most of this work has been done using in-person formats \cite{feng2016designwearable, kamat2022tangibleconstructionkit, vermeersch2019involving, mccosker2019codesigning, lopez2019codesigning, shiri2016bsr, miao2009tactile, metatla2011crossmodal}. In-person design activities often enhance the quality and depth of data collection by enabling the observation of non-verbal cues such as facial expressions and body language. Further, by conducting design activities in-person, the research team may have greater potential to control the conditions of the activity, thereby obtaining more focused, cleaner data.  In contrast, online design programs offer several advantages. First, these programs allow for more flexible recruitment strategies, enabling researchers to reach a more geographically diverse participant pool and increasing inclusion of participants in terms of cultures and abilities \cite{lobe2022a, dpdproject2021korte}. Second, while researchers have less control over the setting for the design activity, these programs allow for data collection from a wider variety of environments, improving the richness of the collected data.

While in-person interactions are challenging to replicate in online settings, researchers have shown that utilizing appropriate collaboration technology (e.g., video conferencing, discussion forums, and chat platforms) can build rapport despite the lack of physical co-presence. For example, a body of literature that emerged from the  COVID-19 pandemic has demonstrated that it is possible to foster strong connections and trust in virtual environments, especially through Community-Based Participatory Research (CBPR), where researchers who are from the pool of participatory design participants are involved \cite{tariq2023lessons}.

Our design process embraces the advantages of both the in-person and online modes of working.  In each of our three design programs, we combined working with a relatively larger contingent of online participants ($N=18, 37, 24$ for the first, second, and third year of the program respectively) with a smaller contingent of in-person participants ($N=6, 2, 3$ for the first, second, and third year of the program respectively).

\subsection{Iterative Design \& Development}
A co-design process with iterations of improvements to the technology itself fosters a sense of responsibility among researchers and co-designers, ensuring that complex ethical and social questions are addressed through each part of the design process \cite{Southwick2021iterative}. Also, ongoing reflection by both the designers and end users is crucial to developing socially responsible technology, as it uncovers the limitations of current designs and optimizes for different points of view by allowing end users to reflect on and potentially reject the design of a technology iteration \cite{sengers2005reflective}. Unique insights gathered from each iteration of a co-design process can foster the development of assistive technologies that are more responsive to user needs, and with longitudinal participation of participants over a set period of time, it can provide specific insights from users' lived experience \cite{kamat2022tangibleconstructionkit}.  In our work, we utilize iterative design both within years of our design program and across multiple design programs.

\subsection{Mobile Navigation Technologies for the Blind}
There is a long history of researchers and engineers developing technology to augment the traditional cane methods of blind travelers \cite{wiener2010foundations}.  This history traces through the introduction of high-tech cane systems that used sonar or lasers to standalone GPS-based navigation devices \cite{ball2008electronic}, and finally to the present day dominance of smartphone apps for navigation \cite{blindsquare, Lazarillo, ahmetovic2016navcog, Goodmaps, yoon2019leveragingar}.  Within the space of mobile apps, specific approaches can be largely broken down into apps that utilize GPS-only for sensing and those that may employ additional sensing mechanisms.  In the GPS-only category, apps have been developed specifically for blind travellers (e.g., Lazarillo \cite{Lazarillo} and BlindSquare \cite{blindsquare}) and mainstream apps are used widely by the Blind community (e.g., Google or Apple Maps).  As the primary purpose of our research is to enable individuals who are blind to navigate with high precision and indoors (key areas where GPS technology falls short), we will focus on solutions that utilize sensing in addition to GPS.

Smartphone apps for high-precision and indoor navigation can be divided into infrastructure-based and infrastructure-free solutions.  Infrastructure-based solutions rely in either installing additional infrastructure (e.g., Bluetooth beacons) or relying on existing infrastructure (e.g., Wifi access points) to serve as landmarks to aid in localization.  Researchers have developed infrastructure-based systems based on a myriad of approaches such as Bluetooth beacons \cite{kunhoth2020indoor, ahmetovic2016navcog}, Wifi fingerprinting \cite{gallagher2012indoor}, and machine-readable signage \cite{navilens}.  Using  infrastructure for localization typically requires collecting sensory measurements of the infrastructure (e.g., the signal strength of Bluetooth Beacons installed in the environment) at a set of known calibration locations within the environment.  Both the establishment of these calibration locations as well as collecting the data can be a time-consuming processes (e.g., as discussed in \cite{ahmetovic2016navcog}).  Infrastructure-free solutions typically rely on using a combination visual and inertial sensing to create a map of the unmodified structure of the environment.  Commercially available systems such as GoodMaps \cite{Goodmaps} have been developed based on this idea.  In contrast to our system, where both mapping and navigation can be done with an off-the-shelf smartphone,  GoodMaps requires a camera and LiDAR system that costs tens of thousands of dollars during the mapping phase. Visual inertial navigation has also been explored in the academic literature \cite{yoon2019leveragingar, yang2024evaluating, fusco2020indoor, crabb2023lightweight}.  However, none of these approaches is suitable for both indoor and outdoor navigation and collaborative mapping and route sharing (as is possible in our system).

\section{Design Process} \label{sec:designprocess}

\begin{figure*}[htbp]
\centering
  \includegraphics[width=\linewidth]{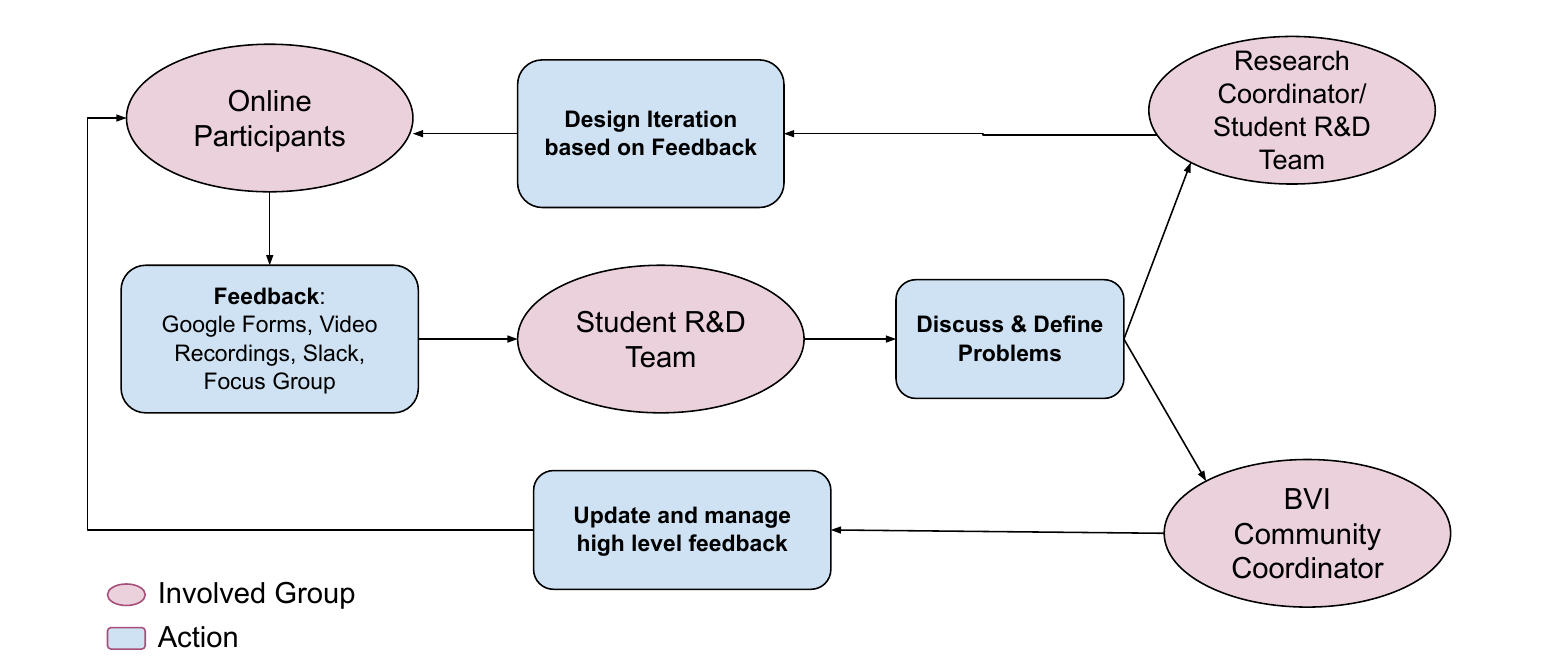}
  \caption{This figure illustrates the feedback design loop. It highlights how online participants, the student R\&D team, BVI community coordinators, and research coordinators contribute to the ongoing development and refinement of the app.}\label{fig:designloop}
  \Description[The feedback process by which our design program operated]{The flow chart shows the various components of the participatory design program.  Online participants provide feedback and design ideas to the R and D team.  The R and D team defines and discusses problems. The BVI community coordinator updates and manages high-level feedback. 
 The research coordinator and the R and D team iterate on the design and implementation of prototypes that further feeds the cycle.}
\end{figure*}

We did not start our design process without prior insights.  Before we began this work, we had collected substantial feedback on our app, \Clew{}, through app reviews on the App Store, Emails, and in-app feedback surveys. We articulated our overarching research questions based on this data analysis.

\begin{enumerate}
    \item How can we design a smartphone application that addresses both indoor and outdoor navigation, allows users to share maps with others, and a provides a reliable and user-friendly positioning system?
    \item How can we collaborate with BVI individuals to iteratively design this smartphone app to be easy to use, easy to learn, fully accessible, and versatile?
\end{enumerate}

To address these questions, we adopted an action research methodology \cite{stringer2020action} to explore and enhance the navigation experience of BVI individuals through the creation of a smartphone app. In our application of this methodology, we aimed to shift the role of the BVI users from subjects to study (i.e., those evaluating already finished technology) to active research partners (i.e., individuals who shaped the early features of technology, sometimes over multiple years). Building on the principles of action research, we took a participatory and co-design approach to understand the preferences and usability challenges of BVI individuals with navigation technology. Our goal was to engage in a process of ``collective-creativity'' \cite{sanders2008co} where ideas, feedback, and lived experiences were freely shared between the participants and the R\&D team.

As mentioned in Section \ref{sec:onlinevsinperson}, involving researchers who are from the community of participatory design participants can help facilitate communication and trust \cite{tariq2023lessons}. We adapted this strategy through the inclusion of BVI individuals as \textbf{co-designers} who directly engaged with and were part of the research and development (R\&D) team.  Further, members of the BVI community worked to manage and streamline the process of the program through the role of program coordinators. These roles were crucial in guiding the team in creating a program that effectively addresses the needs of BVI participants, ensuring the project remains inclusive and responsive to the community. The R\&D team was composed of the principal research coordinator, undergraduate researchers, BVI co-designers and BVI program coordinators. BVI participants who were part of the human-centered, participatory design process, collaborated with the R\&D team, virtually, in the iterative design process. The expectations for the participatory design participants are listed in Table \ref{table:expectations}. The program was conducted during our host institution's undergraduate summer research program, where undergraduate researchers were recruited to join the R\&D team for 10 weeks in the summer.

\begin{table*}[htbp]
    \centering
    \caption{List of expectations given to the participatory design participants throughout the programs.}
    \label{table:expectations}
    \begin{tabular}{ | c | p{0.7\textwidth} | }
    \hline
    \textbf{Expectation} & \textbf{Description} \\
    \hline
    (a) & Generate ideas for features and use cases of an indoor (and outdoor for program year 3) navigation mobile app. \\
    \hline
    (b) & Share experiences with \Clew{}, if any, and other navigation apps and tools. \\
    \hline
    (c) & Test app prototypes during each iteration. \\
    \hline
    (d) & Participate in group ideation sessions with other BVI participants to improve the app usability before and after prototype testing. \\
    \hline
    (e) & Provide feedback through written comments via Slack or Email, one-on-one or focus group virtual calls via Zoom, and/or screen recordings during testing via Slack, Email, or Google Forms. \\
    \hline
    \end{tabular}
\end{table*}

Participants gave feedback consisting of ideas for new features, suggestions on improving existing features, usability or bug reports, insights on use cases for the technology in personal or professional settings, and comments on the organization of the overall design program. In addition, we hosted accessibility-focused career and human-computer interaction (HCI)  workshops with optional but encouraged attendance from the participants.  These workshops were aimed at increasing engagement among the participants, informing them of the resources that are available to them, and helping them better understand and engage with the design approaches used by the R\&D team.

The three annual programs, functioned as a cyclical action research process with four phases. Its goal was to actively enhance the context-specific experiences of BVI individuals through the substantial engagement.

\begin{table*}[htbp]
    \centering
    \caption{Participation requirements and program-specific criteria.}
    \label{table:participation_requirements}
    \begin{tabular}{ | c | p{0.7\textwidth} | }
    \hline
    \textbf{Requirement} & \textbf{Description} \\
    \hline
    (a) & Proficiency with and access to an iOS device. \\
    \hline
    (b) & The ability to conduct design activities in English \\
    \hline
    (c) & Time to attend or engage with program events. \\
    \hline
    (d) & Comfortable mapping and exploring public spaces. \\
    \hline
    \end{tabular}
\end{table*}

\paragraph{\textbf{Planning Phase - Recruitment \& Selection Criteria:}} This phase involved recruiting potential participants through a mixture of channels and coordinating future on-site testing logistics with partner organizations New England College of Optometry (NECO) and the Carroll Center for the Blind. In addition to reaching out to our personal networks, we advertised on public forums like the r/Blind sub-Reddit and the accessible technology site AppleVis. We also emailed the users of {\Clew} to solicit interest in the program. All program participants were compensated for their participation and the study was approved by our Institutional Review Board (IRB). Participants meeting all of requirements (a), (b) and (c) from Table \ref{table:participation_requirements} were selected for program year 1 and 2. In program year 3, we also required (d) from Table \ref{table:participation_requirements} to focus on creating accessible public maps.

\paragraph{\textbf{Acting Phase - Design Process:}} During this phase, participants tested prototypes in real-life scenarios, providing specific insights into the app's practical use. The R\&D team focused on enhancing both the technology and user support, cyclically improving the user interface (UI) and functionality based on real-time feedback. Participants continuously provided feedback and interacted with the app through channels like Slack and virtual meetings.  Additionally, app prototypes enabled participants to record their experiences using the app (consisting of images, narrative feedback, and sensor data), which were utilized by the R\&D team to better understand the feedback from the participants.

\paragraph{\textbf{Observing Phase - Collecting Feedback:}} After each acting phase, the R\&D team gathered qualitative feedback through one-on-one interviews, online focus groups, and Slack communications, identifying the pain points of the app and strengths. We also used exit surveys with open-ended questions and ratings to enhance participant engagement and collaboration \cite{marwaa2023using} and improve the program's organization for future iterations .

\paragraph{\textbf{Reflecting Phase - Reviewing Feedback:}} The reflection phase extended beyond each summer program. At the end of each annual program, the R\&D team and coordinators analyzed feedback to identify improvement areas and shape future agendas. This cyclical research approach (see Figure~\ref{fig:designloop}) deepened our understanding of user experiences and guided ongoing technology improvements.

\begin{table*}[htbp]
    \centering
    \caption{Program participants year-by-year who were members of the BVI community. 
     For online participants, we only included people who were part of the full summer program.}\label{table:participants}
    \begin{tabular}{l | c | c | c |}
    & \multicolumn{3}{c}{\textbf{BVI Participants}} \\
    & Online Participants & R\&D Team Members (Co-Designers) & Program Coordinators \\
    \hline
    \textbf{Year 1} & 18 & 6 & 1 \\
    \textbf{Year 2} & 37 & 2 & 2 \\
    \textbf{Year 3} & 24 & 3 & 3 \\
    \hline
    \textbf{Total Unique} & 67 & 11 & 4
    \end{tabular}
\end{table*}

\subsection{Design Program Year 1: Using \Clew{} as a Design Probe}

\subsubsection{Overview of Design Team}

During year 1 of the program using the aforementioned recruiting strategies, we recruited 18 remote BVI participants (14 from the U.S., where our institute is based, and 4 international) (see Table~\ref{table:participants}). The participants had a range of visual acuity, including individuals with no light perception and those who were low vision but still used vision to perform many daily life tasks. We also collaborated with 6 in-person BVI co-designers in the R\&D team and 1 in-person BVI program coordinator who helped arrange and manage the program (see Table~\ref{table:participants}). All participants met at least one of (a), (b) or (c) requirements listed in Table \ref{table:participation_requirements}. The program was conducted for 5 weeks during the summer.

\begin{table*}[htbp]
    \centering
    \caption{The features of the various applications designed during the duration of the study. 
     The table also provides a timeline of when specific apps were part of the design program.}\label{table:features}
    \begin{tabular}{l | c | c c c |}
    & \textbf{\Clew{}} & \textbf{\ClewMaps{}} & \textbf{\InvisibleMap{}} & \textbf{\PathBlazer{}} \\
    \hline
    \textbf{Route Recording} & yes & yes & no & yes \\
    \textbf{Physical Anchoring} & yes & no & no & no \\
    \textbf{Siri Shortcuts} & yes & no & no & no \\
    \textbf{Navigation Networks} & no & no & yes & yes \\
    \textbf{Visual Localization} & yes & yes & no & yes \\
    \textbf{Fiducial Marker Localization} & no & yes & yes & no \\
    \textbf{Outdoor navigation} & no & no & no & yes \\
    \textbf{In-App Tutorials} & yes & no & no & yes \\
    \textbf{Easily Shareable Wayfinding Data} & no & yes & yes & yes \\
    \hline
    \textbf{Worked on in Year 1} & yes & yes & yes & no \\
    \textbf{Worked on in Year 2} & no & yes & yes & no \\
    \textbf{Worked on in Year 3} & no & no & no & yes 
    \end{tabular}
\end{table*}

\subsubsection{Program Structure \& Process}

We began by using \Clew{} as a design probe to identify the features needed in an indoor navigation app that could help BVI individuals navigate in unfamiliar settings.  Although many participants were familiar with \Clew{}, most did not use the app often or had encountered difficulties that prevented them from fully utilizing its features.  Since many traditional participatory design activities rely on sketching and other visual methods that were not accessible to most of our participants, we explored scenario-based ideation meetings, where participants generated ideas for a navigation app based on their specific use cases and potential circumstance they might encounter. The R\&D team used participants' direct suggestions and concerns to enhance \Clew{} and explore the possible development of a new application that would enable routes to be shared more easily between users (a common request from the participants).

The program started off with semi-structured, one-on-one interviews with one or two members of the R\&D team. The meetings helped to elicit information about each participant's background, experiences with \Clew{}, if any, and strategies for navigating unfamiliar indoor spaces. These initial interviews helped identify pain points with \Clew{}, assess the strengths and weaknesses of current navigation tools, and foster trust with participants.

Following initial interviews, participants tested the App Store version of \Clew{} in their homes or unfamiliar settings during the first week. They then joined focus groups via Zoom to discuss potential UI improvements and suggest new features. Participants used Slack for asynchronous questions and feedback, and Google Forms for screen recordings and feedback submissions.

The R\&D team and BVI participants then identified usability challenges and functional limitations in \Clew{} and aimed to resolve them through the development of two new app prototypes \ClewMaps{} and the \InvisibleMap{}, which we discuss further in Section \ref{sec:designprogram1outcomes}.  Although \Clew{} was originally designed for indoor navigation, many participants hoped for outdoor navigation, particularly during the transition from outdoor to indoor environments (which the R\&D team acted on in design program year 3, see Table~\ref{table:features}).

Based on these suggestions, using an iterative design approach, the R\&D team pursued several core projects (see Table~\ref{table:features}). Online BVI participants would try prototype versions of an idea, provide feedback via online focus group meetings, Slack, Email, or Google Forms and suggest new ideas or concepts to supplement the core idea. Instead of being given formal instructions on testing the prototypes, participants were encouraged to explore the use cases of the prototypes at home and in public indoor places, particularly the unfamiliar ones, to find usability issues or bugs, generate ideas for new features, and provide the R\&D team with insights from their own use cases. The R\&D team reached out to the participants via Slack to ask for periodic updates, and scheduled virtual group meetings with them on, approximately, a weekly basis. In the final online feedback meeting, R\&D team members posed high-level questions about the program experience overall.

\begin{table*}[htbp]
    \centering
  \captionsetup{width=0.8\textwidth} 
  \small 
  \centering
  \caption{Co-designer \& Participant Exit Survey Feedback on Open-Ended Questions}
  \label{tab:openendedexitsurveyfeedback}
  \begin{tabular}{cc|*{4}{p{2cm}}}
    \toprule
    & \multicolumn{5}{c}{\textbf{Percentage of Feedback that Mentions the Following}} \\
    \cmidrule(lr){3-6}
    \textbf{Summer} & \textbf{\# of Responses} & \textbf{Enjoyed Collaboration} & \textbf{Liked Involvement} & \textbf{Communication Difficulties} & \textbf{Program Structure Difficulties} \\
    \midrule
    2021 & 11 & 54.55\% & 81.82\% & 45.45\% & 27.27\% \\
    2022 & 35 & 74.29\% & 80.00\% & 45.71\% & 31.43\% \\
    2023 & 22 & 90.91\% & 95.45\% & 13.64\% & 31.82\% \\
    \bottomrule
  \end{tabular}
\end{table*}

\begin{table*}[htbp]
    \centering
  \captionsetup{width=0.8\textwidth} 
  \small 
  \centering
  \caption{Co-designer \& Participant Exit Survey Feedback on the Ratings of Program \& Value of the Application}
  \label{tab:exitsurveyratingsfeedback}
  \begin{tabular}{cc|ccc}
    \toprule
    & \multicolumn{4}{c}{\textbf{Ratings from 1-10 where 1 is no value and 10 is great value}} \\
    \cmidrule(lr){3-5}
    \textbf{Summer} & \textbf{\# of Responses} & \textbf{Impact \&}  & \textbf{Re-participate in}  & \textbf{Value of App \&}  \\
    & & \textbf{Contribution} &\textbf{Program} & \textbf{Likely to Continue Using App} \\
    \midrule
    2021 & 11 & 8.31 & 9.62 & 8.92 \\
    2022 & 35 & 6.59 & 9.27 & 8.94 \\
    2023 & 22 & 7.64 & 9.68 & 8.91 \\
    \bottomrule
  \end{tabular}
\end{table*}

\begin{table*}[htbp]
    \centering
  \captionsetup{width=0.8\textwidth} 
  \small 
  \centering
  \caption{Categorization of Co-designers' \& Participants' Relevant Slack Messages}
  \label{table:slackmessagecategorization}
  \begin{tabular}{lcccccc}
    \toprule
     & \textbf{\# of Messages} & \textbf{Help/Question} & \textbf{Logistics} & \textbf{App Feedback} & \textbf{Response to Others} & \textbf{Idea for a New App} \\
    \midrule
    \textbf{2021} & 76 & 14.5\% & 15.8\% & 42.1\% & 18.4\% & 9.2\% \\
    \textbf{2022} & 56 & 25.0\% & 21.4\% & 8.9\%  & 42.9\% & 1.8\% \\
    \textbf{2023} & 81 & 37.0\% & 0.0\% & 48.1\% & 11.1\% & 3.7\% \\
    \bottomrule
  \end{tabular}
\end{table*}

\subsubsection{Outcome: App Features} \label{sec:designprogram1outcomes}

\paragraph{\textbf{1. Interactive Tutorials: }} 
The R\&D team and BVI participants identified challenges regarding user interface inconsistencies with \Clew{} and noted the steep learning curve that many participants faced when using \Clew{} for the first time. One particular point of confusion was the process for loading a saved route to start navigation, which required physically aligning the phone against an orienting surface (e.g., a door-frame or wall).  While the app included a detailed instruction in the app of how to perform this alignment, based on the interview, only a few participants were willing to read the lengthy instruction or understood the instruction. Given these challenges, the team created a set of interactive tutorials to guide users through each specific step (see Table~\ref{table:features}). 

\paragraph{\textbf{2. Siri Shortcuts: }} 
Several participants mentioned a desire to have a hands-free option for controlling \Clew{}, due to the difficulty of operating a phone while holding other items (e.g., a mobility cane). Participants suggested on Slack that \textit{``...the users could create custom commands to make Siri activate these shortcuts...After calling upon a shortcut for a route, I don't think the users should need to press a button to start the navigation''}. In response, the R\&D team implemented Siri shortcuts for recording and navigating routes without requiring physical interaction with the phone screen.

\paragraph{\textbf{3. Cloud-Based System: }} 

The R\&D team implemented the suggestion to create a cloud-based system for sharing routes, resulting in the development of \ClewMaps{}. This application offers significant advantages over \Clew. First, route alignment is achieved by recognizing the position of an image affixed to a wall or doorway, addressing previous issues with physical alignment. Second, NFC tags enable easy route sharing between users. Scanning a tag loads a list of available routes from the cloud, which is then displayed to the user. \ClewMaps{} reached a mature stage of development by the end of program year 1.

\paragraph{\textbf{4. Developing \InvisibleMap{} with April Tags for Connected Paths: }} 

\begin{figure*}
  \includegraphics[height=2in]{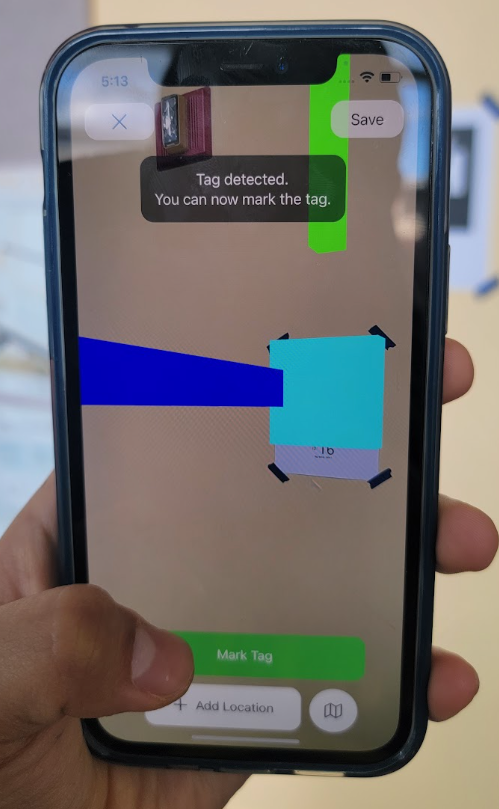}
  \includegraphics[height=2in]{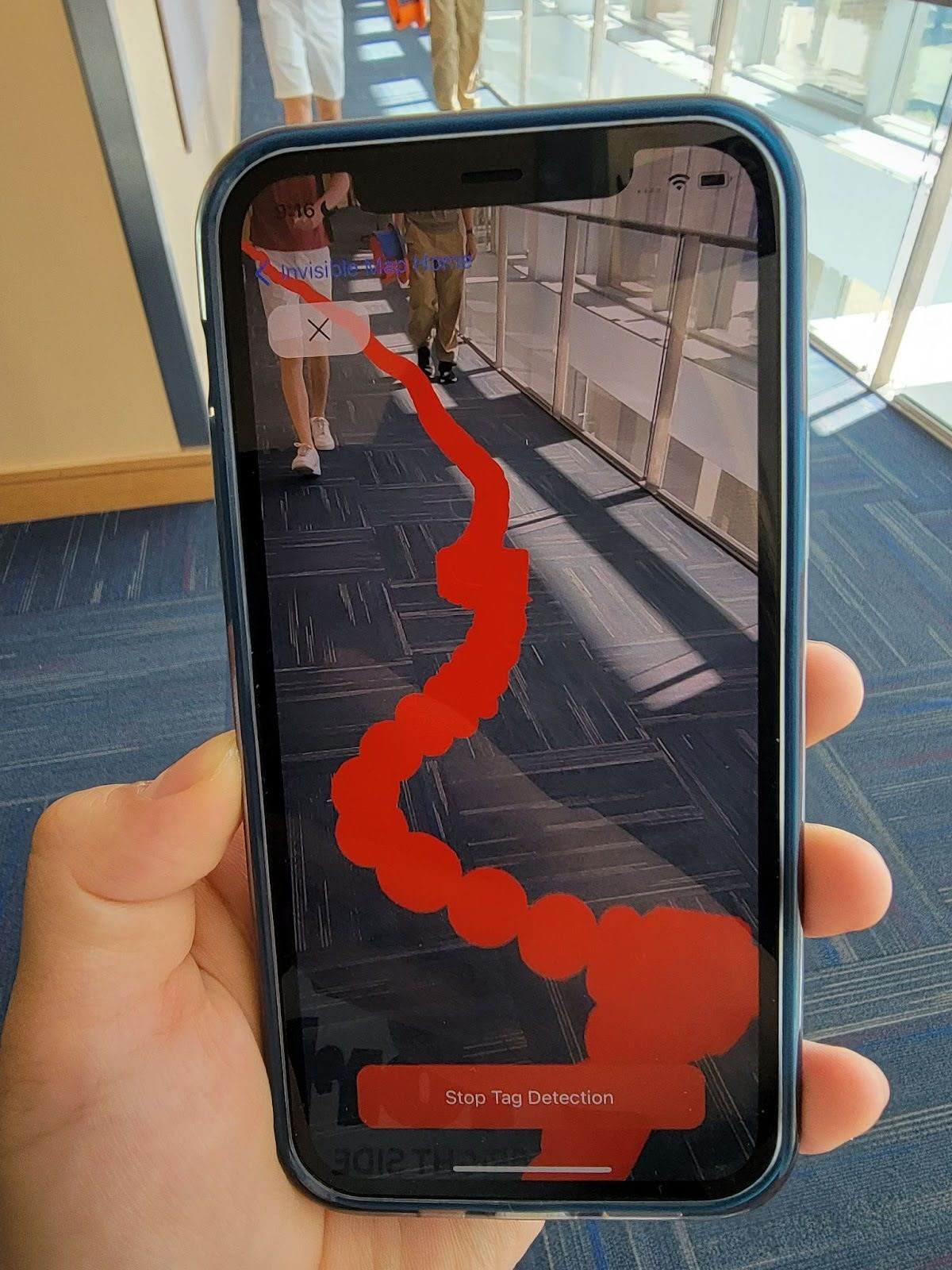}
  \caption{\textbf{Left:} a user marks a detected tag in \InvisibleMap{} in order to add it to the map.  \textbf{Right:} a dynamically generated shortest path route from the user's current position to a goal.}\label{fig:invisiblemap}
  \Description[In one image, a user captures imagery of a fiducial marker to create a map of an environment.  In the second image, a navigation route is shown on a phone screen guiding a user to a destination]{In one image, a user captures imagery of a fiducial marker to create a map of an environment.  In the second image, a navigation route is shown on a phone screen guiding a user to a destination}
\end{figure*}

A final difficulty expressed by BVI participants was the need to record separate routes to connect all pairs of start and end destinations within an environment, since for both \Clew{} and \ClewMaps{}, routes were independent and route segments could not be shared (e.g., to connect points A, B, and C, routes would have to be created between all pairs of points even if large portions of these paths were common).  To move from a route-based to a map-based representation (that would allow sharing of route segments) we developed an app called \InvisibleMap{} to an early stage, with the core functionalities for detecting AprilTags \cite{wang2016apriltag} (fiducial markers that determine location and orientation relative to the camera) and navigation based on shortest path algorithms. Although the technology and user interface needed improvement, this system offered a network of points, allowing users to navigate between any two points (see Figure~\ref{fig:invisiblemap}).  By using multiple image-based markers for localization (AprilTags), it had improved accuracy over \ClewMaps{} by resetting the phone's odometry drift (its sense of motion relative to an initial starting location) each time a marker was encountered.

\subsubsection{Outcome: Program Reflection}

To improve the design program, we gathered feedback on the program through exit surveys. Participants expressed the need for more organized communication channels within the program. Out of the participants who completed the exit survey, which asked open-ended questions regarding the program (see Table \ref{tab:openendedexitsurveyfeedback}),  45.5\% expressed Slack communication difficulties and a desire to separate various topics into different Slack channels instead of one channel.  27.3\% mentioned that they experienced some confusion with the program's  schedule and expressed a desire to have a structured schedule (like a syllabus) with deadlines and clarification regarding participant expectations at each stage of the program. From the open-ended question, \textit{"What did you enjoy most about the Co-Design program?"}, 54.5\% of the participants mentioned their enjoyment in collaborating with the developers and other BVI individuals, and 81.8\% mentioned that they liked being involved in the development and design process of the app because they felt that their feedback and suggestions were taken into action by the developers or felt that their contribution and feedback was meaningful to the BVI community. Participants provided high ratings on Likert-scale questions regarding their perceived impact and contribution, willingness to re-participate in the program, and likelihood of using \Clew{} (see Table \ref{tab:exitsurveyratingsfeedback}).


To better understand the nature of interactions between participants in the program, we also analyzed the Slack messages to determine their types and relevance to our program structure, aiming to identify key areas for further enhancement (see Table \ref{table:slackmessagecategorization}).  This analysis shows that participants used Slack for a wide variety of functions, and over the three years of the program certain categories of feedback became less prevalent (e.g., ideas for new apps decreased from the first program to the last).

\subsection{Design Program Year 2: Exploring Route Sharing \& Navigation Networks}

\subsubsection{Overview of Design Team}

In the second year, we expanded our design program to 37 BVI participants (28 from the U.S. and 9 international), where 2 were repeating participants from year 1, 2 BVI co-designers in the R\&D team and 2 BVI program coordinators (see Table~\ref{table:participants}). The same participant selection criteria and process were used from the previous year. The program was conducted for 9 weeks during the summer.

\subsubsection{Program Structure \& Process}

We had three main goals for program year 2. First, we aimed to collect feedback on the usability of \ClewMaps{} that implemented route sharing and image-based alignment to the starting point of a route, as discussed in \ref{sec:designprogram1outcomes}. Second, we sought to refine the design of \InvisibleMap{} that aimed to enable users to navigate through a network of points on an indoor map, rather than being restricted to direct navigation between two specific locations, as discussed in \ref{sec:designprogram1outcomes}. Lastly, we aimed to enhance the organization of the program by incorporating feedback from the previous year.

To address usability concerns of the physical alignment of \Clew{} at the start of navigation, as discussed in Section \ref{sec:designprogram1outcomes}, and technological limitations preventing users from sharing recorded routes, \ClewMaps{} was established in year 1 and received usability feedback during year 2 for improvements. Furthermore, to overcome the limitation of uni-directional routes and the inability to share route segments in both \Clew{} and \ClewMaps{}, we continued to focus on the design of \InvisibleMap{}.

With two main ongoing projects, \ClewMaps{} and \InvisibleMap{}, we created separate focus groups for each, and allowed participants to choose which project to participate in, if not both. Both apps require physical images for navigation so the R\&D team distributed design kits for testing. Participants in the \ClewMaps{} focus groups were mailed packages containing NFC tags, paper images for starting anchors, and an instruction manual for testing. Similarly, participants in the \InvisibleMap{} focus groups were provided with AprilTags \cite{wang2016apriltag} and an analogous testing manual. After testing \ClewMaps{} or \InvisibleMap{}, participants provided feedback through focus group sessions, submission forms, and video recordings of app usage similar to year 1.

The methods of communication, interview structures and design iteration structures remained the same as the previous year, but the team moved away from the single-channel communication approach to separate channels on specific topics (e.g., \textit{Announcements}, \textit{\ClewMaps{} Focus Group}, \textit{\InvisibleMap{} Focus group}, etc.) for more organized communication. To maintain participation rates, all virtual meetings were scheduled through Google Calendar, providing automated reminders, supplemented by Slack and Email reminders. Additionally, to accommodate participants from various time zones, multiple focus group sessions were organized at different times of the day.

\subsubsection{Outcome: App Features} \label{sec:designprogram2outcomes}

The R\&D team collected feedback on \ClewMaps{} use cases, though the main focus remained on \InvisibleMap{}. With \InvisibleMap{} in its early design stage, the team concentrated on refining the user interface and advancing the navigation technology based on participant suggestions.

The team also explored refinements to the back-end, localization and mapping system for \InvisibleMap{}. A full discussion of this system is beyond the scope of this paper, but the mapping approach is fully described in \cite{ruvolo2023invisiblemapvisualinertialslam}.  At a high-level, the improvements made to the system included a better noise model for the AprilTag detections, which enabled the system to compute more accurate maps of an environment.

\paragraph{\textbf{1. \InvisibleMap{} User Experience Development:}}

The Invisible Map project consisted of two separate interfaces: the navigator interface and the map creator interface. The navigator app used AprilTags and phone odometry (through ARKit \cite{arkit}, as we do with all of our apps) to create a 3D map of an indoor space with landmark tags and points of interest.  This map allows users to navigate between any location within the map and a set of predetermined destinations (as marked by the user during the mapping process).  The creator app allowed users to create maps of an indoor setting with points of interest, that were then uploaded to the navigator app, where users can navigate an indoor space using that map. We separated these two tasks to two different apps to separate the responsibility of creating maps from the navigation aspect to potentially allow building managers to create maps of their buildings for BVI individuals to use to navigate through that space.

\paragraph{\textbf{2. Discontinuing \InvisibleMap{} Development \& Limitations in \ClewMaps{}:}}
One major concern of the participants was the social barrier of getting establishments to adopt \ClewMaps{} or \InvisibleMap{}, especially with regard to needing physical images for anchors. While this obstacle was known at the outset of designing the app, it was unexpected to hear from the participants regarding their struggles with getting establishments to adopt accessible technologies on their behalf.  A typical line of argument was that physical images will require the cooperation of establishment owners and could be removed. {As stated by a participant, \textit{``Appealing to the sentiments of building owners is a time consuming process and adds an element of emotional labor to the process for blind people.''}} Given this concern and the maturation of approaches to visual mapping that did not require installation of fiducial markers, we decided to shift away from approaches that required installation of physical infrastructure for mapping.

\begin{table*}[htbp]
    \centering
  \captionsetup{width=0.8\textwidth} 
  \small 
  \centering
  \caption{Co-designer \& Participant Exit Survey Feedback on Self-Reported Types of Contribution They Made to the App}
  \label{tab:exitsurveytypeofcontribution}
  \begin{tabular}{lcc|ccc}
    \toprule
     & \textbf{Summer} & \textbf{\# of Responses} & \textbf{New Features} & \textbf{Usability Issues or} & \textbf{Insights on} \\
    &  &  &  & \textbf{Bug Reports} & \textbf{Lived Experience} \\

    \midrule
    & \textbf{2022} & 35 & 57.58\% & 63.64\% & 75.76\% \\
    & \textbf{2023} & 22 & 77.27\% & 86.36\% & 68.18\% \\
    \bottomrule
  \end{tabular}
\end{table*}

\subsubsection{Outcome: Program Reflection}

Similar to year 1, we collected feedback on the program through open-ended exit surveys (see Table \ref{tab:openendedexitsurveyfeedback}). Despite the improvements we made to the  organization of our communication channels, the survey revealed that 45.7\% of respondents experienced communication difficulties related to Slack, including periods of idle time where tasks were unclear due to slow or absent communication from the R\&D team. Although participants' ratings for their likelihood to re-participate and use \ClewMaps{} and \InvisibleMap{} remained high (see Table \ref{tab:exitsurveyratingsfeedback}), there was a decline in their feelings of impact and contribution to the app's research and development.  We attribute this decline to the larger size of the program ($N=37$ versus $N=18$) and the maturation of some of the app concepts and prototypes.


On the positive side, 74.3\% of participants reported that they enjoyed collaborating with developers and other BVI individuals, and 80.0\% felt that their involvement in the development and design process was meaningful, indicating that their contributions were valued by the developers and had a positive impact on the BVI community (see Table \ref{tab:openendedexitsurveyfeedback}). This reflects the success of the program's goals related to community engagement and involvement. To assess the impact of their contributions, participants were asked to select from several options regarding the types of contributions they made (see Table \ref{tab:exitsurveytypeofcontribution}): \textit{Provided suggestions for new features}, \textit{Found usability or other user experience issues and communicated them to the team}, \textit{Provided insights on the experience and needs of people who are blind or visually impaired}. Of the respondents, 57.6\% indicated that they provided suggestions for new features, 63.6\% identified usability or user experience issues and communicated them to the team, and 75.7\% provided insights into the experiences and needs of BVI individuals.



\subsection{Design Program Year 3: Exploring Outdoor and Infrastructure-free Mapping and Navigation}

\subsubsection{Overview of Design Team} \label{sec:program3designteam}
In year 3, we worked with 24 BVI participants, including 19 from the U.S. and 5 international participants. Of these, 3 had participated in both previous programs, 6 were returning from program year 2, and 1 from program year 1. The team also included 3 co-designers and 3 BVI program coordinators (see Table~\ref{table:participants}). We intentionally recruited fewer participants compared to the previous year to provide more individual attention. As discussed previously, we amended the participant selection criteria by requiring (d) from Table \ref{table:participation_requirements} (comfort with mapping and exploring public spaces). The program was conducted for 5 weeks during the summer undergraduate research program of the host institution.

\subsubsection{Program Structure \& Process}

Building on the structure and processes established during year 2, prior to year 3, the R\&D team integrated the best features of the previously developed apps into a new app called \PathBlazer{}, detailed in Section~\ref{sec:finalproduct}. The focus of the program was to gather feedback on the usability of an early-stage prototype of \PathBlazer{} and collaboratively refine the app over the course of the program.

During year 3, the team addressed feedback about ongoing difficulties with the organization of the design program by introducing a more structured approach. Participants received weekly communications outlining their tasks and notifications for scheduled workshops. A dedicated website was created to centralize all necessary materials, including Zoom links and recordings. Weekly workshops covered a range of topics, such as user experience design, career insights, and accessibility, enhancing the participant experience. The program guided participants through weekly app testing, gradually introducing more complex aspects of the application. The R\&D team handled most scheduling and logistics via Email rather than Slack, using calendar invites for meetings and reminders, with rescheduling also managed through Email. This separation expedited the process of addressing feedback on Slack.

\subsubsection{Outcome: App Features} \label{sec: app3featureoutcomes}

\paragraph{\textbf{1. Mapping and Localization Technology Refinement: }} 

One major focus of year 3 was the refinement and evaluation of the mapping and localization technology for \PathBlazer{}. Since navigation applications are highly dependent on the environment, understanding the outcomes of any design trial requires a thorough analysis of real-world conditions. However, without being physically present to observe and control these environments, the R\&D team had to rely on indirect methods, such as interviews, logging, screen recordings, and other tools. Participants recorded their experiences and ideas as they were interacting with prototypes through the built-in screen recording feature of the phone, a custom voice-recorder feature the team created for users to narrate their impressions and ideas, and a detailed logging system whereby imagery and sensory information from a given design session could be analyzed in detail by the R\&D team to diagnose errors and better validate our mapping and localization technology. By analyzing the motion logs captured by the phone, the R\&D team was able to enhance the precision of the algorithm and optimize the app's fusion of indoor and outdoor maps (e.g., when navigating routes that contained both indoor and outdoor route portions).

\paragraph{\textbf{2. Outdoor Navigation Integration: }} 
To support outdoor navigation, the team incorporated Google StreetView positioning technology, which allows determining a  more accurate outdoor position than with GPS alone through image-based localization. \PathBlazer{} is the first application from this study that supports outdoor navigation. To test this nascent feature, the team gathered feedback by adding points of interest based on the latitude and longitude of locations relevant to each user, such as the entrance to their dentist’s office or their apartment complex's management office. The team then added support for marker-less mapping and localization by integrating Google Cloud Anchors, which use visual SLAM (Simultaneous Localization and Mapping) to convert 30 seconds of visual and inertial data to a map that can be used for relocalization when a user wants to navigate through an environment.

\paragraph{\textbf{3. Streamlining Anchor Creation and Connection Workflow: }} 
To support networks of navigation starting and end-points (as we had previously supported in \InvisibleMap{}), participants helped design a system whereby users can  progressively map the area around navigation starting and end points and then, at a later time, connect these points with route segments.  Further, participants offered suggestions to streamline this process by designing special workflows that  enable the creation of a new navigation destination and a path connecting it to the existing map within a single workflow (rather than through two separate and time-consuming steps).

\paragraph{\textbf{4. Timed Voice Prompts: }}
To help BVI participants better map the local area around a navigation starting or end point, the R\&D team added timed voice prompts that gave instructions as to what sorts of motions would create the best possible anchor (e.g., \textit{``perform a 360-degree sweep with your phone''}, \textit{``take a step back and do a second sweep''}). This change made the creation of high quality anchors accessible to BVI mappers.


\subsubsection{Outcome: Program Reflection}

The refinements to the program resulted in fewer participants expressing difficulty in communicating with the R\&D team and with the program's overall organization. Among the 22 participants who responded to the program's exit survey (see Table \ref{tab:openendedexitsurveyfeedback}), only 13.6\% mentioned that they had communication difficulties, while around 31.8\% stated that they experienced program structure difficulties and suggested adding clearer deadlines for testing iterations and more consistent calendar reminders. Approximately 90.9\% of them mentioned that they enjoyed collaborating with other BVI people and the R\&D team, while 95.5\% mentioned that their involvement in the projects felt impactful.

With more organization and fewer participants, the average score for perceived impact on the app increased from the previous year with participants' willingness to re-participate in the program and use of \PathBlazer{} similarly high as the previous years (see Table \ref{tab:exitsurveyratingsfeedback}). As for the types of contributions they made, 77.3\% indicated that they provided ideas for new features, 86.4\% stated that they provided usability issues to the R\&D team, and 68.2\% stated that they provided insights from their lived experiences as a BVI individual as reflected in Table \ref{tab:exitsurveytypeofcontribution}. 


With scheduling and logistical matters handled via Email, Slack messages predominantly focused on feedback or questions related to the app (see Table \ref{table:slackmessagecategorization}). Some of those messages included, \textit{``When creating an anchor, at least on an iPhone 12 pro screen, the notes field is not selectable and will not switch or bring up the keyboard.''} and \textit{``I created two anchors, connected them, and improved the connection...Now when I went back to the app both of the anchors are gone. Do I have to be near the location for anchors to be listed or are they just gone?''}. In contrast to program year 2, there were less messages revolving around concerns regarding missed Slack notifications and difficulties accessing meeting links, indicating an improvement to the organization of the program structure. 


\section{Final Product: NaviShare}\label{sec:finalproduct}
The culmination of our design process, was the creation of the NaviShare app.  As the previous sections have described, the three key features that NaviShare provides that were not part of \Clew{}, are:
\begin{itemize}
\item Creating maps, across multiple sessions consisting of interconnected networks of starting and end points.
\item Navigating in both indoor and outdoor spaces with no disruption to the users' navigation experience as they change from one type of environment to the next.
\item The ability to share maps between multiple devices (via the cloud) so that maps created by one user can be used by another.
\end{itemize}

For our mapping and positioning algorithm, we built on top of two cloud-based mapping services, Google's Cloud Anchors \cite{cloudanchors} and Google's Geospatial Anchors \cite{geospatial}.  The user experience for both mapping and navigation is designed to be fully accessible for users regardless of whether they have typical vision or are part of the BVI community.

\subsection{Indoor Mapping}\label{sec:indoormapping}

Our mapping system consists of multiple, local maps that are stitched together across multiple mapping sessions.  Each map consists of both anchors (points for starting or ending a navigation route) and connections (paths that connect two anchors).

\textit{\textbf{Anchor Creation.}} Users create anchors using visual inertial odometry (VIO) and simultaneous localization and mapping (SLAM), which creates a point cloud of sparse, visually identifiable landmarks (e.g., interest points) with corresponding coarse 3D positions (as determined using structure from motion techniques).  We do not implement the  underlying SLAM system from scratch, but instead rely on Google's Cloud Anchor system \cite{cloudanchors}, which uses motion estimates derived from Apple's ARKit library \cite{arkit} with a proprietary visual SLAM algorithm to record a map of a space given 30-seconds of visual-inertial data.  While the specific details of Google's visual SLAM algorithm are not available, the documentation of the library and inferences we can make based on commonly used structure-from-motion-based mapping techniques makes it likely that the best mapping performance is attained by slowly scanning through a large portion of the space while translating (not just rotating) the phone.  We coach the user to make a good anchor using verbal instructions that are made accessible using technologies like screen-readers and refreshable Braille displays.  When creating the anchor, we specify a reference pose (position and orientation) that will be used during re-localization.  The specific choice of reference pose is arbitrary (as long as the choice is consistent and known), so we choose a pose located at the origin and with no rotation with respect to the ARKit tracking coordinate system.  Once the anchor has been created, the API provides an identifier that can be used to localize the anchor (and consequently the pose of the origin) within a new tracking session.

\textit{\textbf{Anchor Connections.}} As mentioned above, the position of each anchor is defined relative to the coordinate system of the ARKit tracking session that was active during the anchor creation process.  As such, without an additional connection step, the relative positions of the two anchors (assuming they were connected at different times, and consequently using different ARKit coordinate systems) are unknown.  In the connection phase between anchors \emph{A} and \emph{B}, we seek to solve two challenges.  First, we would like to know the relative transformation between the two anchors.  Second, we want to record a navigable path that can be used to walk between \emph{A} and \emph{B}.

To connect two anchors, the user first selects the starting anchor (which we'll refer to as anchor \emph{A}) and then the ending anchor (which we'll call anchor \emph{B}).  Once the starting and end point are selected the following steps are performed:

\begin{enumerate}
    \item  The user localizes to anchor \emph{A} by panning their camera around the space.  Once the anchor localizes, we move to the next step.
    \item The user walks from \emph{A} to \emph{B} along the path they'd like to use for navigation between those two points.  During this phase, the user keeps the phone's rear camera pointing in the direction of travel.  During this phase NaviShare records the phone's pose to record a trail of virtual breadcrumbs.  Additionally, we use the cloud anchor API to create a \emph{path anchor} to serve as a visual-spatial map of the path itself.  This additional map helps the performance of the system over longer routes.
    \item As the user approaches \emph{B}, the anchor will localize (often several meters before arrival).  At this point the user can continue walking until they reach \emph{B}.  At this point they can hit the stop record button, and the connection has been fully created.
    \item (Optional) If the user has not recorded the path from \emph{B} to \emph{A} (i.e., the reverse path), the app will ask the user if they'd like to record the path in this direction.  Although not required, users can customize the reverse path to match the specific route they would take when walking in the opposite direction (e.g., staying to the right side of a hallway). Additionally, the reverse path will capture visual imagery to establish a path anchor for the reverse direction, enhancing the likelihood of successful localization when traveling that way.
\end{enumerate}

The result of this process consists of visual-spatial maps along the path itself, a navigable set of points within the environment (virtual breadcrumbs), and an estimate of the relative position of anchors \emph{A} and \emph{B} (available since both are now localized within the same ARKit tracking session).

\textit{\textbf{Streamlined Process.}} As discussed previously, in response to feedback from the program participants, we created a process whereby a connection and a new anchor can be created in a single step.  First, the user selects the starting anchor, then they localize to it.  Next, they walk to the location where they'd like to create a new anchor.  Finally, they map the area to create an anchor to mark the end of the connection.  This workflow reduces the number of steps needed to create two anchors and connect them.

\subsection{Outdoor Mapping}

The workflow described in Section~\ref{sec:indoormapping} can be applied outside as well.  That said, we can significantly improve the robustness of our map by referencing each anchor to a precise latitude, longitude, yaw, and altitude.  By determining the precise position and orientation of an anchor with respect to these geo-spatial parameters, we enable users to localize and navigate to anchors from arbitrary starting locations using either as-the-crow flies navigation or utilizing walking direction routing systems like GoogleMaps.  Given our focus on precise outdoor localization and navigation, we cannot simply use GPS and magnetometer data to establish our position and orientation on the Earth's surface (as these are inaccurate for our use case).  We rely on Google's Geospatial API \cite{geospatial}, which fuses camera-based positioning with GPS and inertial data to precisely determine the user's location and orientation.  The camera-based system relies on Google StreetView data, so the accuracy of the geo-spatial position is only high when the user is in an area that has up-to-date StreetView imagery.  Further, it is important that the area has sufficient lighting to enable the system to match the camera images to the StreetView data.  In our testing, we found that the system works well in the day-time and also at night-time in urban areas that have adequate street lighting.

In order to support outdoor mapping, we created a function in our app that allows the user to record an \emph{outdoor anchor} (distinguished from the \emph{indoor anchors} described previously).  When recording an outdoor anchor we use the same four-step process described in section ~\ref{sec:indoormapping} with the addition of an initial step of geo-localization.  During the geo-localization step, we use the Geospatial API to determine the user's latitude, longitude, altitude, and heading.  The Geospatial API provides us with an estimate of these four parameters, which define the user's position and orientation, along with confidence intervals for each.  When the confidence intervals have shrunk sufficiently to indicate a precise localization, we record the user's geo-spatial pose and move to step 1 of the indoor mapping workflow.

\begin{figure}
    \includegraphics[width=0.5\linewidth]{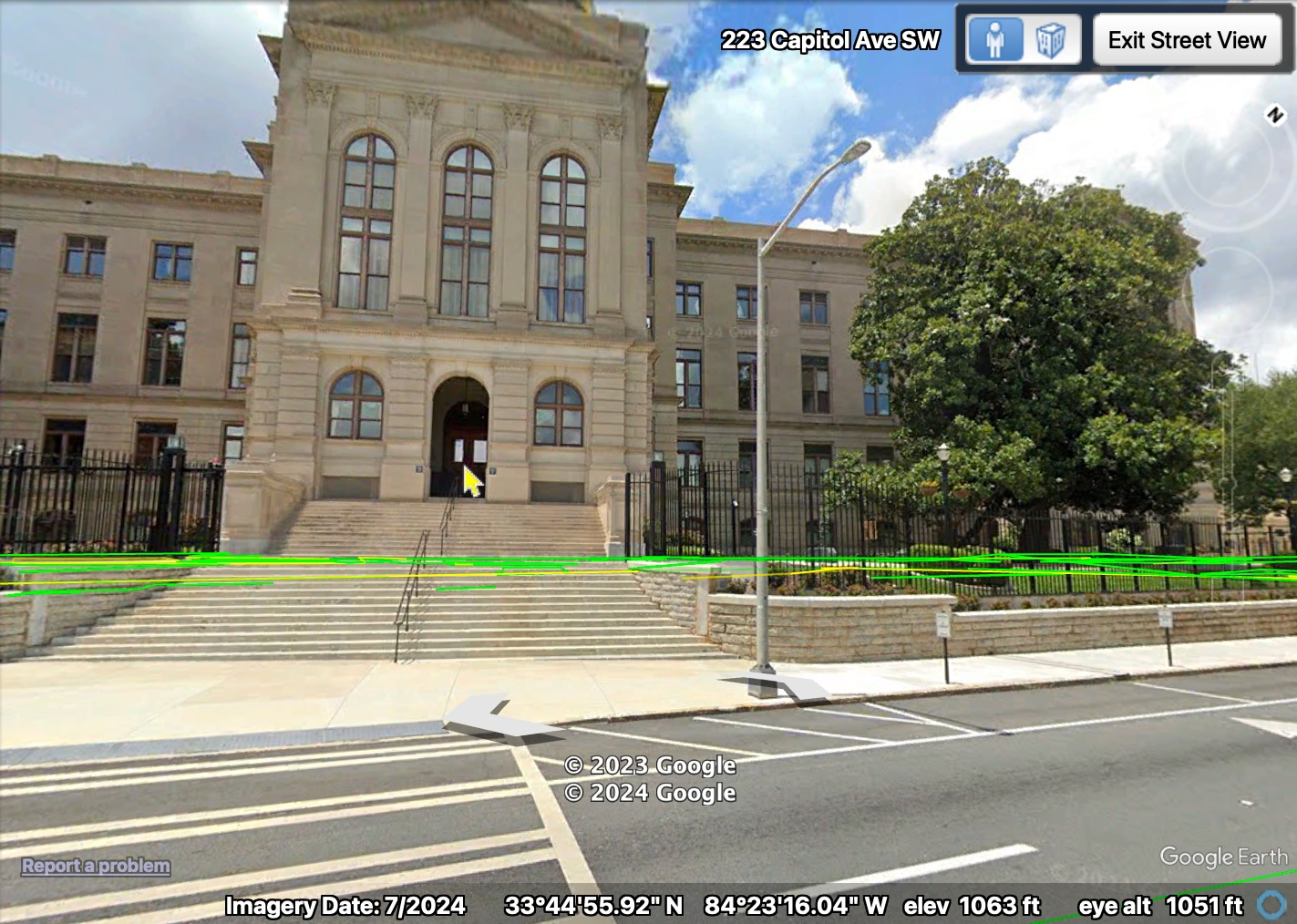}
    \caption{An example of determining the latitude and longitude of a point of interest using Google Earth.  The cursor, shown in yellow, is placed at the point of interest and the latitude and longitude can be read off the info bar at the bottom of the screen.}\label{fig:googleearth}
    \Description[The image shows a church rendered in Google Earth.  A cursor, shown in yellow, is positioned at the door of the church.  The bottom of the image shows information such as the date of the StreetView imagery, the latitude, the longitude, and the altitude.]{The image shows a church rendered in Google Earth.  A cursor, shown in yellow, is positioned at the door of the church.  The bottom of the image shows information such as the date of the StreetView imagery, the latitude, the longitude, and the altitude.}
\end{figure}

We also support a second, virtual outdoor-mapping workflow where the geo-spatial position of a point of interest is determined using street view data annotated through a laptop (i.e., not in situ with a smartphone).  Using Google Earth, we are able to hover our mouse over a street view image of the point of interest and then record the longitude and latitude (see Figure~\ref{fig:googleearth}).  


\textit{\textbf{Navigation.}} The user can explore potential navigation destinations through the app's menus.  The list of possible destinations are filtered based on the coarse location of the user (as determined by GPS).  Once the user selects their preferred destination, a list of possible starting points for navigation will be displayed.  The user then selects their current location at the start of the route.  While choosing one's starting location is not as seamless an interaction as having the phone automatically localize the user, we found that giving the system the approximate starting location helps with system robustness and reduces the occurrence of false localization.  


\textit{\textbf{Routing.}} When the user selects a start and end point for their route, the app computes the shortest path of travel.  We solve this using classical graph planning algorithms (specifically Dijkstra's algorithm).  The topology of our graph is constructed by considering the anchors as nodes and the connections between them as edges.  We use the length of any connecting path as the edge weight (cost).  As such, our routing algorithm is capable of stitching together multiple paths into longer routes that more efficiently cover complex spaces than our previous app.

To convert each individual route segment into a unified coordinate system, we use cloud anchors as reference points for alignment. Specifically, since the cloud anchor marking the end of one route segment is also the cloud anchor marking the beginning of the next, we can leverage these shared landmarks to re-align the trail of breadcrumbs (the path connecting two cloud anchors) within a consistent coordinate system.

\textit{\textbf{Positioning.}} After the user selects the start and end points for the route and the route segments are determined, the app assists in precisely localizing the user within their environment. During this localization phase, the user pans their phone back and forth to align the camera feed images with the mapping anchors.

\textit{\textbf{Navigation User Experience.}} Our app uses a combination of haptic, audio, and speech-based cues to guide the user to their destination.  The overall route between the start and end points is divided into segments. As the user navigates these segments, they receive cues that guide them to key points marking the end of each segment. The app provides an audio cue to indicate that the user is on track, meaning if they continue in their current direction, they will reach the key point. If the user veers off track, they can press the \textit{`Get Directions'} button for guidance back to the key point. As the user nears the key point, the intensity of haptic feedback increases, signaling that a directional change may be necessary soon. Upon reaching the key point, the app communicates the next direction along with the distance to travel (e.g., turn left and proceed 20 meters).

\begin{figure*}
\includegraphics[height=2in]{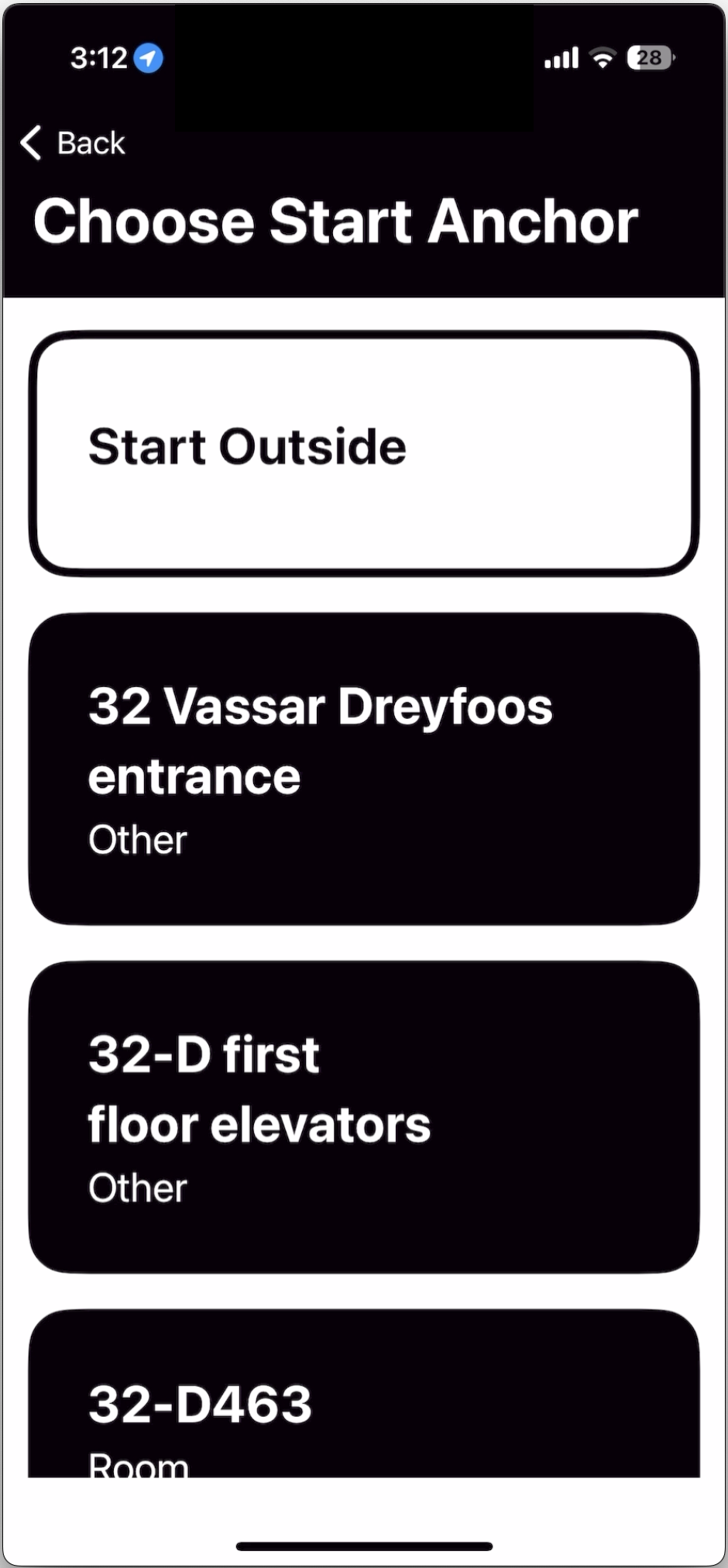}
\includegraphics[height=2in]{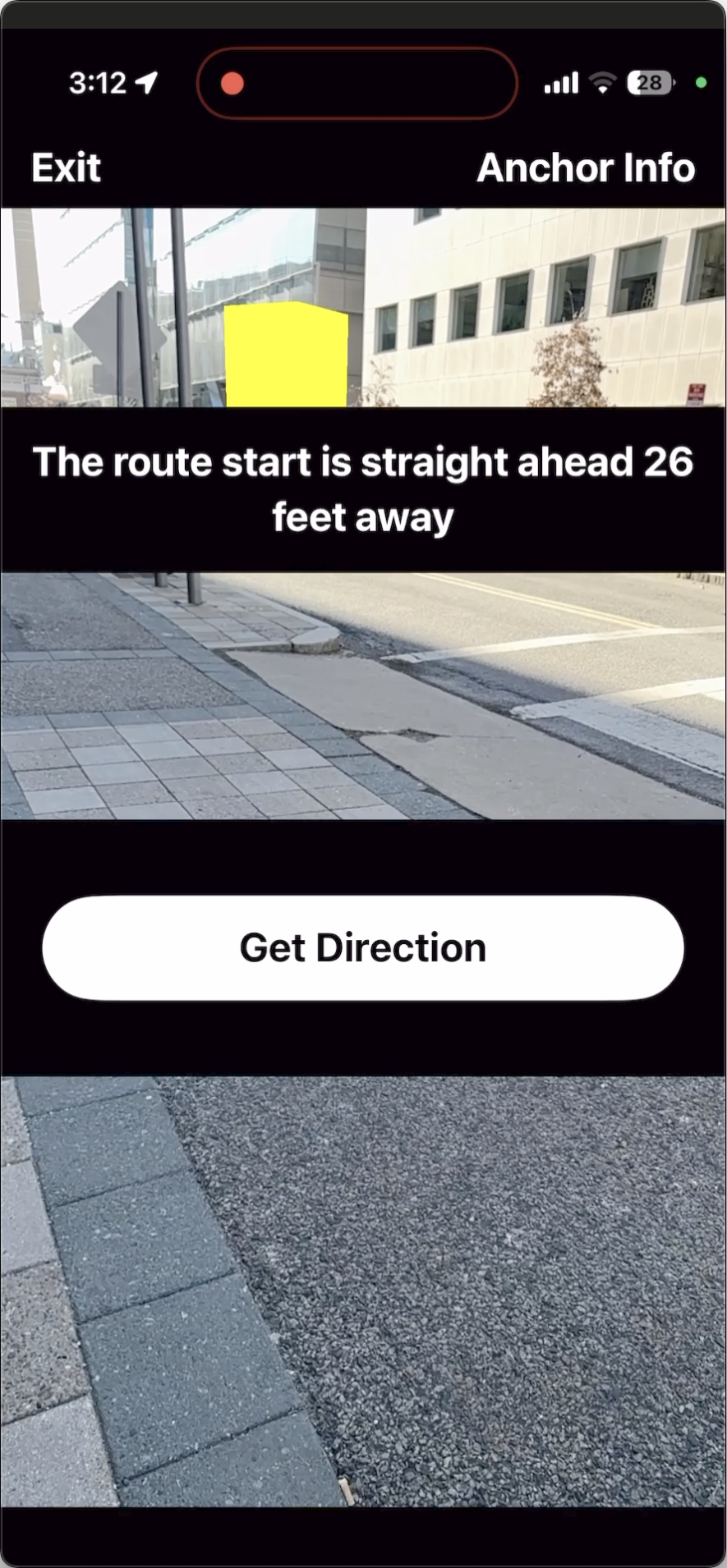}
\includegraphics[height=2in]{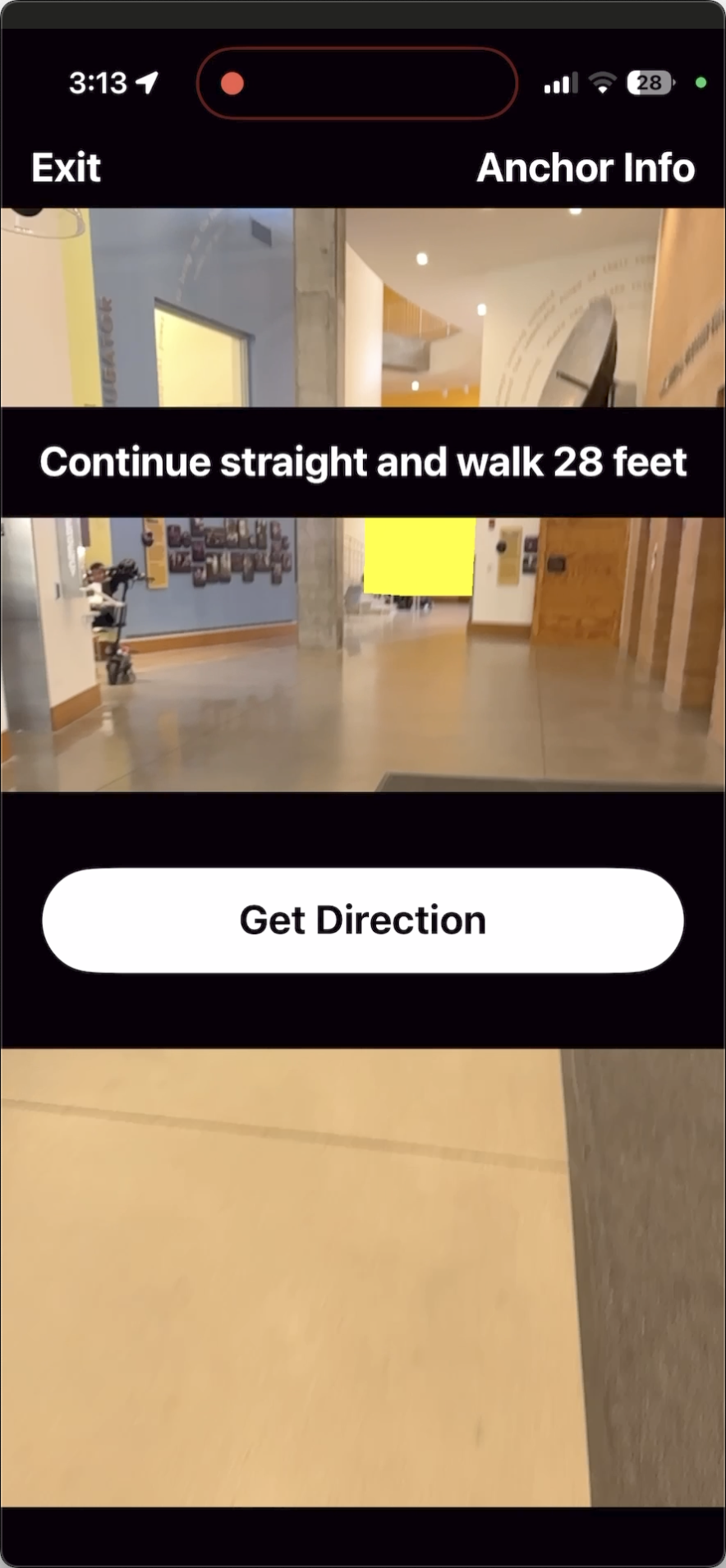}
 \caption{\textbf{Left:} after the user chooses a destination point inside a nearby building, the \PathBlazer{} app presents a list of potential starting locations. \textbf{Center:} after selecting ``Start Outside'', the app determines the user's position using GPS and camera data and begins providing outdoor guidance. 
 \textbf{Right:} as the user enters the building, the app beings using indoor features to refine the user's position and continues providing guidance along the route.}\label{fig:navishare}
 \Description[the left image shows several possible starting locations including starting outside, a street entrance and an elevator.  The middle image shows a direction to the user that the route start is straight ahead 26 feet away.  The camera feed is also shown, which shows an urban landscape with a yellow, virtual beacon superimposed above the start.  The right image shows the camera feed from an indoor portion of the route.  The app is directing the user to continue straight and walk 28 feet.  A yellow, virtual beacon is superimposed to shows the next point along the route.]{the left image shows several possible starting locations including starting outside, a street entrance and an elevator.  The middle image shows a direction to the user that the route start is straight ahead 26 feet away.  The camera feed is also shown, which shows an urban landscape with a yellow, virtual beacon superimposed above the start.  The right image shows the camera feed from an indoor portion of the route.  The app is directing the user to continue straight and walk 28 feet.  A yellow, virtual beacon is superimposed to shows the next point along the route.} \label{fig:navisharescreenshots}
\end{figure*}

\section{Summative Evaluation of \PathBlazer{}}

In addition to the feedback we collected in the design program, we performed a summative evaluation of the \PathBlazer{} app using a more traditional HCI paradigm of an in-person, lab-based assessment.  Specifically, participants used \PathBlazer{} to navigate through a medium-sized building on a university campus to a variety of destinations.  While all participants had been to the building before, the evaluation involved them traveling along parts of the building that they had never been to.  During the evaluation, participants were instructed to select particular starting and endpoints for navigation.  Once selected, the research team observed the participants using the app and analyzed their ability to use the app for successful navigation.

\subsection{Participants}

In total, $N=7$ individuals participated in this study (6 male and 1 female).  All participants met the threshold of legal blindness with a median best-corrected visual acuity score of 20/200 (the minimum acuity was 20/640 and the maximum 20/125) as assessed by one of the study authors who is a board-certified optometrist.  The study was approved by our IRB, and all participants were compensated for their participation.

\subsection{Observation of Outcomes}

The participants were able to use the app to navigate through all routes (2-3 routes were tried per participant).  In some cases, the participants asked for assistance with the app (e.g., getting updated directions if they had overshot a turn or localizing their phone to the environment).

\subsection{Feedback}

We asked the participants to describe their overall experience using the \PathBlazer{} app.  While we did not specifically prompt them to provide a quantitative judgment, some of the participants did so:
\begin{itemize}
    \item P1: ``Intellectually a 5 out of 5, 100\% very positive.  There's nothing else like it.  It was ``enabling;'' \ldots the data at intersections in the hallway, corrections after the forced mistake were very helpful \ldots [I] can see the value overall \ldots I'm emotional because it is so helpful''
    \item P2: "I liked it"; tactile and auditory feedback were best features; fairly intuitive.
    \item P3: ``excellent''
    \item P4: ``pleasant, would definitely use it; would like to be able to pick [destination] beforehand / use GPS''
    \item P5: ``very interesting; 10/10''
    \item P6: ``very positive''
    \item P7: ``8.5 out of 10''
\end{itemize}

We also included a survey question regarding how safe individuals felt using \PathBlazer{} and participants gave an average rating of 4.71 (where 5 is the safest).  This is notable since many of the participants expressed less confidence with both indoor navigation (3.5 average, where 5 is totally confident and 1 is totally unconfident) and using smartphone technology (4.28, where 5 is totally confident and 1 is totally unconfident).

\section{Discussion and Future Work}

Here, we provide a discussion about our design process as well as the most promising future directions for the development of the \PathBlazer{} app.

\subsection{Feedback Mechanisms}

We found that employing multiple, well-defined communication and feedback collection methods allowed us to gather a larger volume of valuable insights from co-designers and participants in real-time, both online and offline.  This diverse methodology was effective partly because it provided co-designers and participants with the flexibility to choose the communication method they were most comfortable with, whether it be focus group or one-on-one virtual meetings, Email, Slack, Google Forms, or in-app feedback (implemented in program year 3). The benefits were evident when collecting asynchronous feedback via Google Forms, which included an option for screen recordings provided during play-testing. This allowed participants to provide detailed, context-rich feedback at their convenience.

Another key benefit of using diverse feedback channels is the facilitation of real-time communication. Slack channels facilitate the rapid exchange of ideas and immediate feedback between the R\&D team and participants. Additionally, one-on-one virtual meetings were conducted to assist participants with play-tests, ensuring that any issues could be addressed promptly. Focus group sessions, typically comprising 5-10 participants and co-designers and led by at least one R\&D team member, were structured to ensure equitable participation.

\subsection{Participant Engagement}

The reflection phases of each design program iteration, which involved analyzing exit survey feedback and Slack messages from participants, provided valuable insights for improving program organization. Acting on different forms of feedback helped reduce communication and logistical difficulties. As a result, participants were able to concentrate on giving valuable feedback on the use cases of the app, rather than spending time navigating resources. 

Despite the limitations of remote participation, weekly group focus meetings proved highly valuable for the app design. BVI participants collaborated on scenario-based questions related to the app’s use cases, exchanging ideas and generating new concepts. However, instances of participants who dominated discussions in these virtual group meetings occasionally occurred, which might have been mitigated by hosting smaller focus groups, facilitating a more balanced exchange of ideas and ensure that all voices were heard \cite{williams2015what}. 

We discovered that providing educational sessions, such as the workshops we offered on accessibility-focused careers, human-computer interaction strategies, relevant technologies, and methodologies, may contribute to positive participant engagement. For instance, one participant expressed their enjoyment of the program by their response to the question \textit{“What did you enjoy most about the Co-Design program?”}, where they answered, \textit{“helping BVI, workshops/focus groups”}. We believe these sessions both allowed participants to gain a better understanding of the technology used in the project, thereby encouraging them to provide informed feedback and ideas, but also fostered a sense of motivation and purpose. This aligns with statement in \cite{laitano2017participatory} that participants are more likely to accept challenges when they perceive the issues addressed are personally meaningful. 

\subsection{In-Person and Remote Feedback Types}

In addition to the main insights from \cite{lobe2022a} on the benefits of online versus in-person collaboration (discussed in Section \ref{sec:onlinevsinperson}), our work reaffirmed that offering the program both online and in-person optimizes its effectiveness for participants, co-designers, and researchers. This dual approach enabled us to include diverse perspectives from different levels of visual impairment and cultural backgrounds through the online component, while also allowing for in-depth, firsthand observation of end-users' challenges during in-person engagements with BVI co-designers.

The integration of online and in-person modalities proved to be complementary, each compensated the limitations of the other. The online option facilitated the engagement of BVI individuals in the design program without the limitations imposed by physical travel. Conversely, the in-person component provided opportunities for direct interactions and nuanced observation of non-verbal cues, which was crucial in understanding users' challenges in real-world contexts. These face-to-face interactions facilitated immediate identification of potential problems, clarification of issues, and a more empathetic design approach.  Ultimately, the integration of both online and in-person modes allowed the strengths of each to compensate for the limitations of the other, leading to more effective and user-centered design outcomes.

\subsection{Future Work on \PathBlazer{}}


Although there were significant improvements to our navigation mobile app over the course of the programs, \PathBlazer{} has some limitations that we aim to respond to, including: (1) the ability for the app to localize itself within large, indoor environments without the user having to pre-specify their approximate location.  This task is difficult due to the repetitive appearance that exists in many indoor environments; (2) participants expressed interest in a fully on-device localization system, removing the requirement to send data to Google's Cloud Anchor API for processing. We are currently working on creating our own, edge-compute pipeline for localization based on \cite{brachmann2023accelerated}; (3) we are working on a more streamlined tool (based on other work \cite{saha2019project}) to make the virtual outdoor-mapping workflow more efficient, so that we can apply to it to the creation of large datasets of outdoor points of interest (e.g., the doors to buildings or bus stops).

\section{Conclusion}\label{sec:conclusion}

This work employed a large-scale, longitudinal, hybrid participatory and co-design program involving BVI individuals to develop indoor and outdoor navigation technology. It introduces \PathBlazer{}, a mobile navigation app for BVI users. Key features of \PathBlazer{} include: (1) the ability to create maps across multiple sessions with connected starting and end points; (2) seamless navigation in both indoor and outdoor environments; (3) the capability to share maps across multiple devices, enabling users to access maps created by others. Over three annual summer programs, we refined a mobile application by incorporating user feedback into each design iteration. We explored various navigation technologies, including physical and virtual markers for alignment and navigation, route sharing among users, and the integration of indoor and outdoor navigation. In design program year 3, we consolidated these features into a single application. The collaboration of BVI participatory design participants, BVI co-designers in the R\&D team, and BVI program leads during the design process not only refined the app's functionality but also enhanced its user interface to accommodate a diverse range of use cases.  This app was evaluated using an in-person, HCI research paradigm at a healthcare facility to validate the designs.



While our programs effectively incorporated participants' preferences for mobile navigation technologies, \PathBlazer{}'s user experience could be enhanced by eliminating the need to specify a starting location and by providing robust, adaptive localization across diverse environments, including complex urban settings, rural areas, and crowded public spaces. Further research on mobile app UI preferences tailored to the BVI community would not only reduce the learning curve for users but also improve accessibility, usability, and customization to meet individual needs. Finally, it is crucial to continuously engage BVI users in the app design process. Their direct involvement ensures that the technologies developed are aligned with their real-world needs, ultimately leading to more inclusive and practical navigation technologies.

\newpage

\bibliographystyle{ACM-Reference-Format}
\bibliography{ref}

\appendix

\end{document}